\documentclass[conference]{IEEEtran}
\usepackage{soul}

\usepackage{cite}

\ifCLASSINFOpdf
   \usepackage[pdftex]{graphicx}
\else
   \usepackage[dvips]{graphicx}
\fi
\usepackage{cmath}
\usepackage{amsmath,amssymb,amsfonts}
\usepackage[absolute]{textpos}
\begin{document}

\begin{textblock}{13}(1.5,0.2)
{\footnotesize \noindent U. De Silva, S. Pulipati, S. B. Venkatakrishnan, S. Bhardwaj, and A. Madanayake, ``A Passive STAR Microwave Circuit for 1-3 GHz  Self-Interference Cancellation," \textit{MWSCAS 2020: 63rd IEEE International
Midwest Symposium on Circuits and Systems}, Springfield, MA, USA, August 2020, pp. 1-4.}
\end{textblock}
%
\title{A Passive STAR Microwave Circuit for 1-3 GHz  Self-Interference Cancellation }

\author{\IEEEauthorblockN{Udara De Silva, Sravan Pulipati, Satheesh Bojja Venkatakrishnan, Shubhendu Bhardwaj and Arjuna Madanayake}
\IEEEauthorblockA{Florida International University (FIU), Miami, USA, Contact: amadanay@fiu.edu}}


%


\maketitle

\begin{abstract}
Simultaneous transmit and receive (STAR) allows full-duplex operation of a radio, which leads to doubled capacity for a given bandwidth. A circulator with high-isolation between transmit and receive ports, and low-loss from the antenna to receive port is typically required for achieving STAR. Conventional circulators do not offer wideband performance. Although wideband circulators have been proposed using parametric, switched delay-line/capacitor, and N-path filter techniques using custom integrated circuits, these magnet-free devices have non-linearity, noise, aliasing, and switching noise injection issues. In this paper, a STAR front-end based on passive linear microwave circuit is proposed. Here,  a dummy antenna located inside a miniature RF-silent absorption chamber allows circulator-free STAR using simple COTS components. The proposed approach is highly-linear, free from noise, does not require switching or parametric modulation circuits, and has virtually unlimited bandwidth only set by the performance of COTS passive microwave components. The trade-off is relatively large size of the miniature RF-shielded chamber, making this suitable for base-station side applications. Preliminary results show the measured performance of Tx/Rx isolation between 25-60 dB in the 1.0-3.0 GHz range, and 50-60 dB for the 2.4-2.7 GHz range.
\end{abstract}

\begin{IEEEkeywords}
full duplex, simultaneous transmit and receive (STAR), self-inteference cancellation (SIC).
\end{IEEEkeywords}

%
\IEEEpeerreviewmaketitle

\section{Introduction}
The advent of 5G standard has galvanized the research on advanced radio systems in both sub 6 GHz frequencies\cite{sub_6_1} and millimeter wave frequencies\cite{comcas_2019, iwat_2019}. In this age of maxed-out spectrum in legacy frequency  bands ($<$6 GHz), increasing demands for capacity on wireless networks has led to several potential solutions.  Among the proposed approaches for capacity increase, notions of dynamic spectrum access in cognitive and full-duplex radios\cite{sharma_2018}, in both the legacy and the mm-wave bands, are perhaps the most well developed. To achieve in-band full-duplex capability, the transceivers are required to achieve 90+ dB isolation between transmitter and receiver\cite{sateesh_1}. This can potentially be accomplished via multiple isolation stages, such as antenna isolation, RF isolation, and digital isolation\cite{sateesh_2, sateesh_3}. In this paper, we propose a novel passive RF isolation circuit that can be used in wideband full-duplex radios along with other antenna isolation and digital isolation techniques.

In full-duplex radios based on a single transmit (Tx)/receive (Rx) antenna, the usual approach is to use a circulator that provides isolation between the transmitter,  being the output port of the power amplifier (PA), and the receiver, which is the input port of the low noise amplifier (LNA). The circulator is a non-reciprocal three-port device. The scattering matrix of a mismatched circulator is shown in Eqn~\eqref{eq:circ_s} where $|\gamma | \ll 1$ represents small imperfections like antenna mismatch and LNA/PA impedance variations. When all three ports are matched (i.e., reflection coefficient $\gamma = 0$), it has perfect isolation from the transmitter to antenna making it suitable for STAR applications. However, when there is a mismatch, the circulator isolation is limited by $\gamma$\cite{pozar_microwave}. Antenna tunning units (ATU) can somewhat mitigate this limitation\cite{garcia_2004}. 

\begin{equation}\label{eq:circ_s}
S_{circulator} = 
\left[ {\begin{array}{*{20}{c}}
\gamma&\gamma&1-\gamma^2\\
1-\gamma^2&\gamma&\gamma\\
\gamma&1-\gamma^2&\gamma
\end{array}} \right],~\gamma\in\mathbb{C}
\end{equation}

An ideal wideband circulator ($\gamma=0)$ isolates the Tx and Rx paths over a wide band of frequencies. Practical circulators are far from ideal in their characteristics, and cause a variety of problems: frequency-dependent responses and mismatch between the antenna, LNA and PA modules, leakage from the PA to the LNA which may cause undesirable self-interference and saturation of the receiver block or a non-linear response of the receiver chain. The circulator may be large in size and generally, also exhibit a narrowband performance when it uses traditional ferrite-based topology. Active integrated circuit based circulators have been proposed that offer relatively wide bandwidths and reasonable isolation, especially on mobile units/handsets.


\section{Wideband integrated circulators}

New types of magnetless circulators based on switched delay lines \cite{ethan_nature_2017,ethan_RWS_2018,ethan_IMS_MMIC_2018}, parametric modulation \cite{alu_circulator_2016,alu_MTT_2017,alu_MTT_2019} and N-path filtering\cite{harish_2019, harish_nature_2016,harish_2017_LPTV} have been proposed and demonstrated. These circulators are compatible with microwave monolithic integrated circuit(MMIC) technology, which allows them to be integrated into transceiver chips. The experimental results have reported 20-50 dB Tx-Rx isolation and 1.7-8 dB Ant-Rx insertion loss, as reported in Table~\ref{tbl:comp}. Magnet-free circulator approaches are indeed promising but nevertheless suffer several detriments; the performance in parametric circuits suffer from narrow bandwidth, high attenuation in the Tx-Ant path, non-linear effects due to modulating of varactors, and poor noise figure due to addition of active components. Switched delay line based circuits have issues due to injection of switching noise, phase-mismatch between the delay lines, frequency-dependent behavior, and relatively low band of operation within the primary Nyquist zone. 
%
%

The drive to create wideband circulators for STAR is based on the fact that the circulator is a classical microwave component and is an obviously good approach for achieving full-duplex STAR operation. However, \emph{circulation is sufficient but not necessary} for full-duplex STAR operation. In fact, circulators are over-specified for this application.  Unidirectional coupling from Tx-Ant ports is not necessary for STAR, nor is unidirectional coupling between the Rx-Tx ports. Therefore, we can find alternatives based on non-circulation approaches that lead to the desired STAR full-duplex functionality. One ideal example is a 3-dB splitter/combiner, where the antenna is connected to the sum port and transmitter and receiver on the remaining ports. If there is perfect isolation and perfect antenna matching over a wide band, a circulator can be replaced with a splitter at the cost of a 3-dB insertion loss to the received signal.

\begin{figure}[t!]
\centering
\includegraphics[scale=0.19]{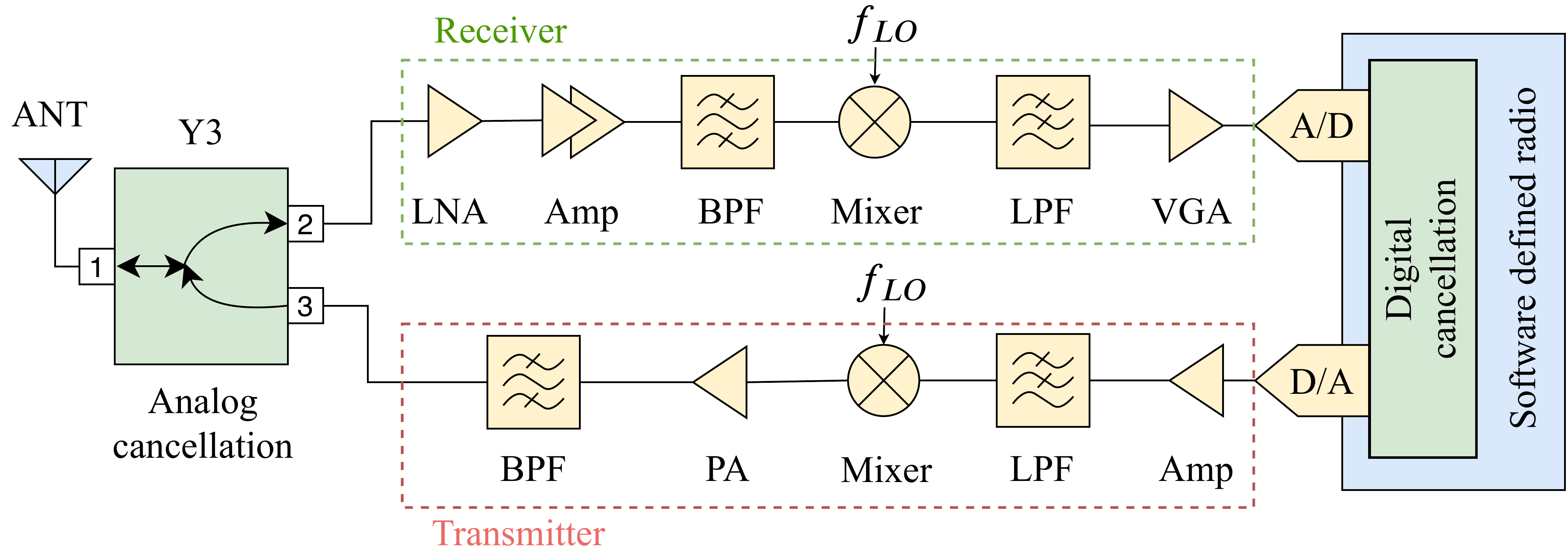}
\vspace{-1ex}
\caption{A typical full duplex system architecture which shares the same antenna for transmit and receive, by introducing the RF block Y3.}
\vskip -3ex
\label{fig:system}
\end{figure}
\section{Active Interference Canceller}

Instead of relying on the standard circulators only, there were attempts to build analog interference canceller (AIC) circuits using active circuits\cite{ASIC_1,ASIC_2,ASIC_3}. In one approach, the self-interference signal is modeled by analog tapped delay line (ATDL) filters that have weights which are either trained by digital assistance\cite{digital_AIC_2017} or optimized by analog feedback loops\cite{alms_AIC_2017}. Another approach known as electrically balanced duplexing (EBD) tries to develop tunable LC circuits to imitate the impedance of the antenna\cite{Haine_COMM_2015}. Active analog interference canceller circuits can achieve over 50 dB Tx-Rx isolation\cite{EBD_ISSCC_2015}. While the small size and ability to fabricate on-chip makes AIC circuits more desirable for hand-held devices, they are limited by the bandwidth, complexity, noise/distortion, and power consumption. To support wide-bandwidth AIC circuits approaches yield high order filters that suffer from  circuit complexity, power consumption, and long tuning/calibration times. We propose  a passive self-interference canceller that trades physical size for very wideband performance, ultra low noise, zero distortion, zero power consumption (except for losses), while not requiring control circuits and/or calibration.

\section{Passive Interference Canceller: Twin Antenna Circuits}
A STAR system is shown in Fig.~\ref{fig:system}, as is needed to utilize an antenna for both transmission and reception for full duplex radios. Therefore, to isolate the transmit and receive chains, a 3-port RF module Y3 (which is shown using a green box in Fig.~\ref{fig:system}) is required, such that a relatively large transmit signal does not present itself at the LNAs input.  RF channel paths between the ports 2 to 1 and the ports 1 to 3 should exhibit small losses. 

\begin{figure}
\centering
\includegraphics[scale=0.22]{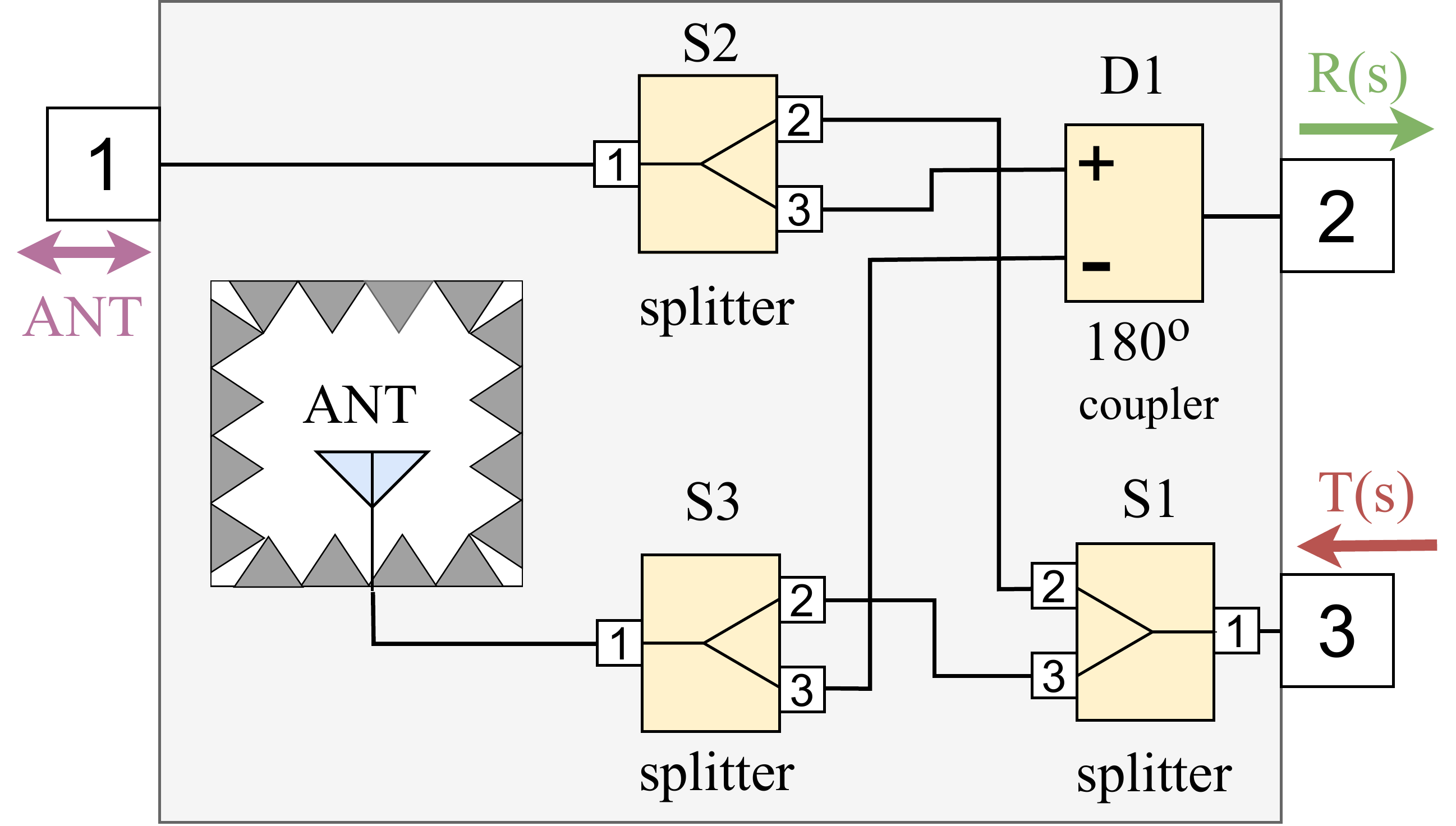}
\vspace{-1ex}
\caption{Y3: The proposed full-duplex wideband linear isolator circuit.}
\vskip -3ex
\label{fig:sci_circuit}
\end{figure}

We propose a self-interference cancellation scheme where an identical antenna that resides in a miniature RF shielded box is used in a symmetrical parallel connection to the Tx/Rx communication antenna.
The proposed twin-antenna system relies on creating a copy-of and then subtracting the coupled transmit signal from the received signal. Therefore, we refer to this scheme as twin-antenna wideband canceler. A schematic showing the proposed functionality of the Y3 block is shown in Fig. \ref{fig:sci_circuit}. All the cables in the design are phased matched to minimize the phase mismatches. Effectively, the proposed assembly of three splitters and a dummy antenna is used to create a copy of the signal, which could couple from Tx to Rx chain. The created copy is then subtracted from the received signal, using the difference block $D_1$ leaving only the desired received signal. 

The copy of the coupled signal is created by performing the following operations. First, the transmit signal $T(s)$ is split equally between topologically symmetric arms, using splitter $S1$, resulting in forward split transmit signal components $H_{s,f}(s)T(s)$ in each of the arms, where $H_{s,f}(s)$ is the forward transfer function of the splitter. We assume that the splitter is well balanced, and the isolation between the output ports 2 and 3 is high. The split transmit signal enters $S2$ and appears at the Tx/Rx communication antenna at the bidirectional port 1. The split transmit signal also enters $S3$ and appears at the Tx/Rx dummy antenna. Because the dummy antenna is identical to the communication antenna, the driving point impedance $Z_A(s)$ is the same as that of the communication antenna, given that both the antennas are fabricated with tightly controlled tolerances (that is, twin-antennas). The frequency-dependent transmit power coupling from ports 2 to 3 in $S2$ and $S3$ are identical. The dummy antenna should not receive any received signals from the air interface; therefore, it is placed inside a miniature RF shielded box to guarantee both reflection-free and RF-silent operation. Therefore, the residual leakage due to splitter non-ideality at port 3 of $S2$ and port 3 of $S3$ are both $T(s)H_{s,f}(s)H_{s,L}(s)$.

Under ideal conditions, desired component of the received RF signal could be obtained from subtraction to produce a zero self-interference. 
The interference-free received signal at the output port 3 of Y3 can be derived as follows. The communication antenna inserts the received signal $R(s)$ on the $S2$, making the combined signal at port 3 to be $R(s)H_{s,f}(s)+T(s)H_{f,s}(s)H_{s,L}(s)$. Since dummy antenna is located in an RF-silent region, the signal received at the port 3 of the $S3$ is only the one coupled from the transmitter, i.e., $T(s)H_{s,f}H_{s,L}(s)$. Therefore,  the output of the $180^o$ coupler becomes $(R(s)H_{s,f}(s)+T(s)H_{f,s}(s)H_{s,L}(s)-T(s)H_{s,f}H_{s,L}(s))H_{h,f}(s)$ where
$H_{h,f}(s)$ is the forward function of the $180^o$ coupler. The output is therefore, ideally, given by $R(s)H_{s,f}(s)H_{h,f}(s)$. Given, $180^o$ coupler and splitter are sufficiently wideband and have approximate anticipated transfer function $H_{s,f}(s)=0.707$ and $H_{h,f}(s)=1$, the output signal is given by $0.707R(s)$. This implies a 3-dB loss in the proposed configuration, while the receiver does not experience the transmit signal $T(s)$ at the input port of the LNA, as one desired in a full-duplex single antenna scheme. The proposed system also isolates the receiver chain from reflected power from port 2 of splitter S2, S3, and power amplifier due to the symmetry of the circuit.

The key idea is the use of an identical antenna inside the miniature RF-shielded box. In regular use, RF absorbers are used in the far-field, which is $\gg \lambda$, where $\lambda$ is the wavelength of the lowest frequency radiated. However, in our experiments, we found out that a miniaturized RF-shielded box with dimensions of $\approx\lambda$ does not significantly alter the reflection coefficient of the antenna ($S_{11}$). This observation leads to the implementation of twin-antenna cancellation systems with practical size for base stations. In our prototype design which operates in 1-3 GHz frequency range (i.e., $ 30~cm \geq \lambda \geq 10~cm $), the RF-shielded box has dimensions of $20\times 24 \times 35~cm^3$. For higher frequencies, the dimensions of the RF-shielded box is further reduced (volume scales as $V\approx\lambda_{Max}^3~m^3$). The flat RF absorber sheets\cite{MAST_sheets} used in the RF-shielded box provide 45 dB loss per one-inch thickness. Fig.~\ref{fig:vna_measurement} shows how the dummy antenna follows the wideband impedance curve of the communication antenna, while exactly modeling all the scattering present due to the communication antenna and the splitter $S1.$ In effect, the dummy antenna acts like an analog computing block that models the coupling from Tx to Rx chain in the presence of the communications antenna. This modeling would be challenging to achieve using a finite order tapped delay-line finite impulse response filter (FIR) or other lumped circuit because of the necessary high order transfer functions make such a solution impractical. By using an exact copy of the communication antenna where the dummy antenna is needed, we avoid the difficult tapped-delay FIR filter requirement and obtain an exact solution ``essentially for free'', the trade-off being the size and the associated 3dB power loss. 
\vspace{-2mm}

\section{Example Implementation}
\begin{figure}
\centering
\includegraphics[width=0.46 \textwidth]{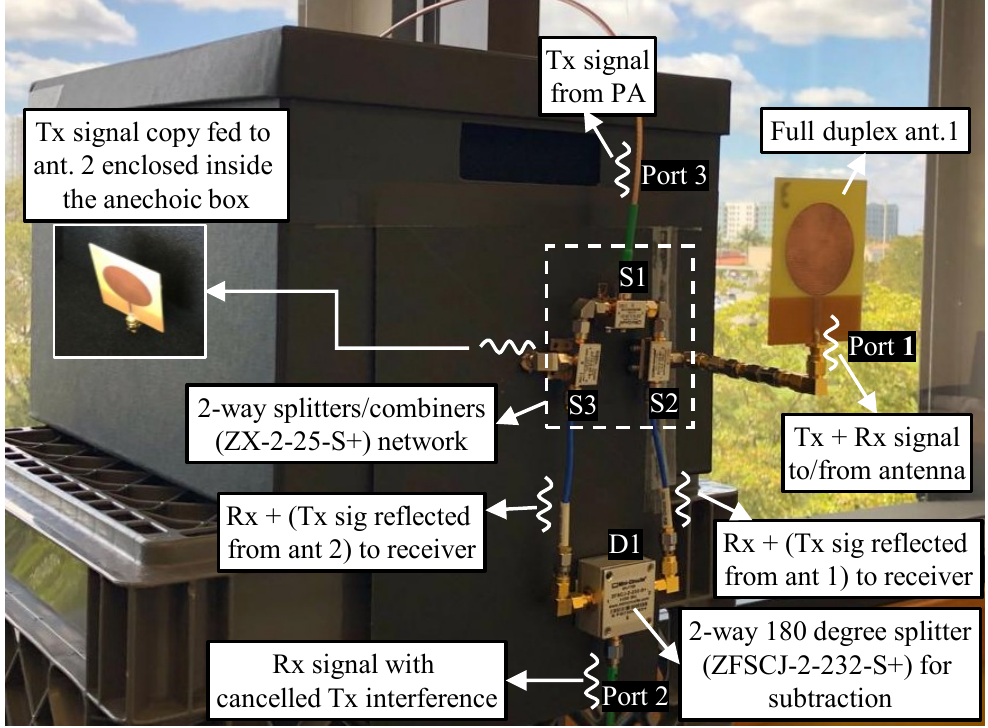}
\vskip -1ex
\caption{Prototype for the proposed full duplex wideband linear isolator circuit.}
\label{fig:prototype_design}
\end{figure}
To verify functionality of the $Y3$ block, experiments were conducted in the 1-3 GHz range. The $Y3$'s realization is shown in Fig.~\ref{fig:prototype_design}. We designed $Y3$ using low-cost commercial off-the-shelf (COTS) components. The Minicircuits part ZX-2-25-S+ is used for the splitters and ZFSCJ-2-232-S+ for the subtractor $D1$. The splitters have 20 dB of isolation; we expect the performance of the $Y3$ to improve with better (and likely a lot more expensive) power splitters and $180^o$ couplers having improved bandwidth and isolation, but the versions used in this paper are meant only as an initial proof of concept implementation. 

%
\begin{figure}
\vskip -3ex
\centering
\includegraphics[width=0.48 \textwidth]{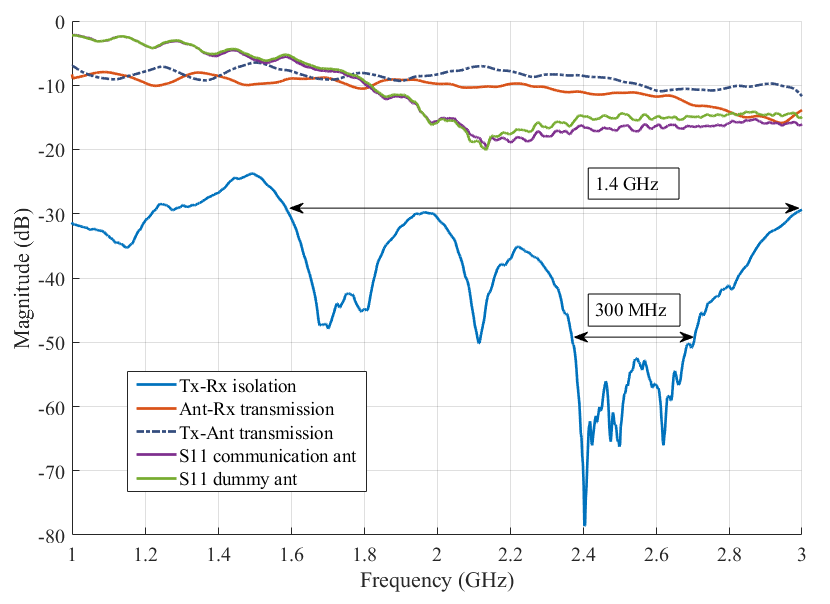}
\vskip -2ex
\caption{VNA Measurements: Tx-Rx isolation (blue), Ant-Rx transmission (red), and Tx-Ant transmission (dotted blue) of the proposed full duplex circuit. The purple and green curves show the measured reflection coefficient of the communication antenna and the dummy antenna respectively.}
\vskip -4ex
\label{fig:vna_measurement}
\end{figure}
%
%
%




The Tx/Rx port isolation (Port 2 to Port 3) is measured using a vector network analyzer (VNA) ( Part number: NI PxIE 5632) and results are shown in Fig.~\ref{fig:vna_measurement}. The measured isolation is the best-case performance of $Y3$ component \emph{in the case the loss from antenna to Rx is zero valued}. However, we know that there is a 3 dB loss of the received signal due to the splitter at port 1 of $Y3$. Microwave components, such as the connectors, cables, and the $180^o$ coupler contain their own losses. Losses reduce the received signal power available at the input of the LNA, and raise the noise temperature of the receiver. The absolute performance of $Y3$, measured using  the VNA shows Tx/Rx isolation between 25-60 dB for the 1.0-3.0 GHz range, between 30-60 dB in the 1.6-3.0 GHz range, and better than 50 dB for the 2.4-2.7 GHz range. These measurements imply that $Y3$ has 6.5-33 dB improvement in Tx/Rx isolation when compared to the performance from using a splitter, which has a nominal performance of about 20 dB. The measured values are compared with previous work in Table~\ref{tbl:comp}.
\vskip -2ex
\begin{table}[h!]
\centering
\caption{Full-duplex performance comparison with previous work. TA: Twin Antenna, NPF: N-path filtering, SSDL: Sequentially switched delay line, STCM: Spatiotemporal conductivity modulation, EBD: Electrically balanced duplexing. }\label{tbl:comp}
\begin{tabular}{|l|c|c|c|c|}
\hline
Device                 & \begin{tabular}[c]{@{}c@{}}Bandwidth \\ (GHz)\end{tabular} & \begin{tabular}[c]{@{}c@{}}Tx-Rx isolation \\ (dB)\end{tabular} & \begin{tabular}[c]{@{}c@{}}Ant-Rx IL\\   (dB)\end{tabular} & Technology \\ \hline
Proposed               & 1.0-3.0                                                          & 25-60                                                           & 8-15                                                       & TA        \\ 
                       & 1.6-3.0                                                        & 30-60                                                           & 8-15                 &                                    
                       \\ 
                       & 2.4-2.7                                                        & 50-60                                                           & 11-13                                                     
                       
                        &           \\ \hline
\cite{harish_nature_2016}   & 0.45-1                                                     & 40-50                                                           & 1.7-3                                                      & NPF        \\ \hline
\cite{harish_SiRF_2019}     & 0.735-0.765                                                & 15-50                                                           & 1.7                                                        & NPF        \\ \hline
\cite{ethan_IMS_MMIC_2018} & 1-2                                                        & 20-35                                                           & 5-8                                                        & SSDL       \\ \hline
\cite{campbel_2019}          & 1.15-1.2                                                   & 22-25                                                           & 3-6                                                        & SSDL       \\ \hline
\cite{harish_2017}           & 22.7-27.3                                                  & 18-20                                                           & 3.2-4                                                      & STCM       \\ \hline
\cite{EBD_ISSCC_2015}        & 2.0-2.15                                                  & 55-75                                                           & 3.9                                                        & EBD       \\ \hline
\end{tabular}\\
\vskip -4ex
\end{table}

\section{Conclusion}
A wideband passive self-interference canceller STAR circuit is proposed that can enable full-duplex radios. The experimental results of the prototype design showed over 25 dB Tx/Rx isolation in 1.0-3.0 GHz bandwidth and over 50 dB Tx/Rx isolation in 2.4-2.7 GHz bandwidth. The high insertion loss in the prototype design is due to the use of low-cost COTS components. The splitter and $180^o$ coupler used in the experiment shows $\approx 4$ dB and $\approx$ 5 dB insertion loss, respectively. The rated maximum frequency of the components is 2.5 GHz, which explains the high insertion loss near 3 GHz. These losses can be alleviated by using splitters and couplers that are optimized for the frequency band of interest. The canceller can be extended to build full-duplex wideband MIMO systems, which will be studied in future work.


\bibliographystyle{IEEEtran}
\bibliography{sic_ims}

\begin{thebibliography}{10}
\providecommand{\url}[1]{#1}
\csname url@samestyle\endcsname
\providecommand{\newblock}{\relax}
\providecommand{\bibinfo}[2]{#2}
\providecommand{\BIBentrySTDinterwordspacing}{\spaceskip=0pt\relax}
\providecommand{\BIBentryALTinterwordstretchfactor}{4}
\providecommand{\BIBentryALTinterwordspacing}{\spaceskip=\fontdimen2\font plus
\BIBentryALTinterwordstretchfactor\fontdimen3\font minus
  \fontdimen4\font\relax}
\providecommand{\BIBforeignlanguage}[2]{{%
\expandafter\ifx\csname l@#1\endcsname\relax
\typeout{** WARNING: IEEEtran.bst: No hyphenation pattern has been}%
\typeout{** loaded for the language `#1'. Using the pattern for}%
\typeout{** the default language instead.}%
\else
\language=\csname l@#1\endcsname
\fi
#2}}
\providecommand{\BIBdecl}{\relax}
\BIBdecl

\bibitem{sub_6_1}
Y.~{Li}, C.~{Sim}, Y.~{Luo}, and G.~{Yang}, ``12-port 5g massive mimo antenna
  array in sub-6ghz mobile handset for lte bands 42/43/46 applications,''
  \emph{IEEE Access}, vol.~6, pp. 344--354, 2018.

\bibitem{comcas_2019}
S.~{Pulipati}, V.~{Ariyarathna}, U.~D. {Silva}, N.~{Akram}, E.~{Alwan},
  A.~{Madanayake}, S.~{Mandal}, and T.~S. {Rappaport}, ``A direct-conversion
  digital beamforming array receiver with 800 mhz channel bandwidth at 28 ghz
  using xilinx rf soc,'' in \emph{2019 IEEE International Conference on
  Microwaves, Antennas, Communications and Electronic Systems (COMCAS)}, 2019,
  pp. 1--5.

\bibitem{iwat_2019}
S.~{Pulipati}, V.~{Ariyarathna}, U.~{De Silva}, N.~{Akram}, E.~{Alwan}, and
  A.~{Madanayake}, ``Design of 28 ghz 64-qam digital receiver,'' in \emph{2019
  International Workshop on Antenna Technology (iWAT)}, 2019, pp. 193--196.

\bibitem{sharma_2018}
S.~K. {Sharma}, T.~E. {Bogale}, L.~B. {Le}, S.~{Chatzinotas}, X.~{Wang}, and
  B.~{Ottersten}, ``{Dynamic Spectrum Sharing in 5G Wireless Networks With
  Full-Duplex Technology: Recent Advances and Research Challenges},''
  \emph{IEEE Communications Surveys Tutorials}, vol.~20, no.~1, pp. 674--707,
  2018.

\bibitem{sateesh_1}
S.~B. {Venkatakrishnan}, A.~{Hovsepian}, E.~A. {Alwan}, and J.~L. {Volakis},
  ``{RF Cancellation of Coupled Transmit Signal and Noise in STAR across 1 GHz
  Bandwidth},'' in \emph{2019 URSI International Symposium on Electromagnetic
  Theory (EMTS)}, 2019, pp. 1--4.

\bibitem{sateesh_2}
S.~{Bojja-Venkatakrishnan}, E.~A. {Alwan}, and J.~L. {Volakis}, ``{Simultaneous
  Transmit and Receive System with 1 GHz RF Cancellation Bandwidth},'' in
  \emph{2018 IEEE International Symposium on Antennas and Propagation USNC/URSI
  National Radio Science Meeting}, 2018, pp. 1241--1242.

\bibitem{sateesh_3}
S.~{Bojja Venkatakrishnan}, E.~A. {Alwan}, and J.~L. {Volakis}, ``{Wideband RF
  Self-Interference Cancellation Circuit for Phased Array Simultaneous Transmit
  and Receive Systems},'' \emph{IEEE Access}, vol.~6, pp. 3425--3432, 2018.

\bibitem{pozar_microwave}
D.~M. Pozar, \emph{{Microwave Engineering}}.\hskip 1em plus 0.5em minus
  0.4em\relax John Wiley \& Sons, Inc., 2012.

\bibitem{garcia_2004}
J.~{de Mingo}, A.~{Valdovinos}, A.~{Crespo}, D.~{Navarro}, and P.~{Garcia},
  ``{An RF electronically controlled impedance tuning network design and its
  application to an antenna input impedance automatic matching system},''
  \emph{IEEE Transactions on Microwave Theory and Techniques}, vol.~52, no.~2,
  pp. 489--497, 2004.

\bibitem{ethan_nature_2017}
M.~Biedka, R.~Zhu, Q.~M. Xu, and Y.~E. Wang, ``{Ultra-Wide Band Non-reciprocity
  through Sequentially-Switched Delay Lines},'' \emph{Scientific
  Reports-Nature}, vol.~7, pp. 1--16, Jan 2017.

\bibitem{ethan_RWS_2018}
M.~{Biedka}, Q.~{Wu}, X.~{Zou}, S.~{Qin}, and Y.~E. {Wang}, ``{Integrated
  time-varying electromagnetic devices for ultra-wide band nonreciprocity},''
  in \emph{2018 IEEE Radio and Wireless Symposium (RWS)}, Jan 2018, pp. 80--83.

\bibitem{ethan_IMS_MMIC_2018}
A.~M. {Darwish}, M.~{Biedka}, K.~{Kingkeo}, J.~{Penn}, E.~{Viveiros}, H.~A.
  {Hung}, and Y.~E. {Wang}, ``{Multi-Octave GaN MMIC Circulator for
  Simultaneous Transmit Receive Applications},'' in \emph{2018 IEEE/MTT-S
  International Microwave Symposium - IMS}, June 2018, pp. 417--419.

\bibitem{alu_circulator_2016}
N.~A. {Estep}, D.~L. {Sounas}, and A.~{Alù}, ``{Magnetless Microwave
  Circulators Based on Spatiotemporally Modulated Rings of Coupled
  Resonators},'' \emph{IEEE Transactions on Microwave Theory and Techniques},
  vol.~64, no.~2, pp. 502--518, Feb 2016.

\bibitem{alu_MTT_2017}
A.~{Kord}, D.~L. {Sounas}, and A.~{Alù}, ``{Differential magnetless circulator
  using modulated bandstop filters},'' in \emph{2017 IEEE MTT-S International
  Microwave Symposium (IMS)}, June 2017, pp. 384--387.

\bibitem{alu_MTT_2019}
A.~{Kord}, M.~{Tymchenko}, D.~L. {Sounas}, H.~{Krishnaswamy}, and A.~{Alù},
  ``{CMOS Integrated Magnetless Circulators Based on Spatiotemporal Modulation
  Angular-Momentum Biasing},'' \emph{IEEE Transactions on Microwave Theory and
  Techniques}, vol.~67, no.~7, July 2019.

\bibitem{harish_2019}
A.~{Nagulu}, T.~{Chen}, G.~{Zussman}, and H.~{Krishnaswamy}, ``{A Single
  Antenna Full-Duplex Radio using a Non-Magnetic, CMOS Circulator with In-Built
  Isolation Tuning},'' in \emph{2019 IEEE International Conference on
  Communications Workshops (ICC Workshops)}, May 2019, pp. 1--6.

\bibitem{harish_nature_2016}
N.~Reiskarimian and H.~Krishnaswamy, ``{Magnetic-free non-reciprocity based on
  staggered commutation},'' \emph{Nature Communications}, 2016.

\bibitem{harish_2017_LPTV}
N.~{Reiskarimian}, J.~{Zhou}, and H.~{Krishnaswamy}, ``{A CMOS Passive LPTV
  Nonmagnetic Circulator and Its Application in a Full-Duplex Receiver},''
  \emph{IEEE Journal of Solid-State Circuits}, vol.~52, no.~5, pp. 1358--1372,
  May 2017.

\bibitem{ASIC_1}
D.~{Liu}, Y.~{Shen}, S.~{Shao}, Y.~{Tang}, and Y.~{Gong}, ``{On the Analog
  Self-Interference Cancellation for Full-Duplex Communications With Imperfect
  Channel State Information},'' \emph{IEEE Access}, vol.~5, 2017.

\bibitem{ASIC_2}
Y.~{Liu}, P.~{Roblin}, X.~{Quan}, W.~{Pan}, S.~{Shao}, and Y.~{Tang}, ``{A
  Full-Duplex Transceiver With Two-Stage Analog Cancellations for Multipath
  Self-Interference},'' \emph{IEEE Transactions on Microwave Theory and
  Techniques}, vol.~65, no.~12, pp. 5263--5273, 2017.

\bibitem{ASIC_3}
J.~G. {McMichael} and K.~E. {Kolodziej}, ``{Optimal tuning of analog
  self-interference cancellers for full-duplex wireless communication},'' in
  \emph{2012 50th Annual Allerton Conference on Communication, Control, and
  Computing (Allerton)}, 2012, pp. 246--251.

\bibitem{digital_AIC_2017}
Y.~{Liu}, X.~{Quan}, W.~{Pan}, and Y.~{Tang}, ``{Digitally Assisted Analog
  Interference Cancellation for In-Band Full-Duplex Radios},'' \emph{IEEE
  Communications Letters}, vol.~21, no.~5, pp. 1079--1082, 2017.

\bibitem{alms_AIC_2017}
X.~{Huang} and Y.~J. {Guo}, ``{Radio Frequency Self-Interference Cancellation
  With Analog Least Mean-Square Loop},'' \emph{IEEE Transactions on Microwave
  Theory and Techniques}, vol.~65, no.~9, pp. 3336--3350, 2017.

\bibitem{Haine_COMM_2015}
L.~{Laughlin}, M.~A. {Beach}, K.~A. {Morris}, and J.~L. {Haine}, ``{Electrical
  balance duplexing for small form factor realization of in-band full
  duplex},'' \emph{IEEE Communications Magazine}, vol.~53, pp. 102--110, 2015.

\bibitem{EBD_ISSCC_2015}
B.~{van Liempd}, B.~{Hershberg}, K.~{Raczkowski}, S.~{Ariumi}, U.~{Karthaus},
  K.~{Bink}, and J.~{Craninckx}, ``{2.2 A +70dBm IIP3 single-ended
  electrical-balance duplexer in 0.18um SOI CMOS},'' in \emph{2015 IEEE
  International Solid-State Circuits Conference(ISSCC) Digest of Technical
  Papers}.

\bibitem{MAST_sheets}
{Mast Technologies}, ``{MF22-0009-00 Lossy Foam Absorber},'' [online]
  Available:
  \url{http://www.masttechnologies.com/products/rf-absorbers/mf2/mf22-0009-00/},
  accessed: 05/18/2020.

\bibitem{harish_SiRF_2019}
N.~{Reiskarimian} and H.~{Krishnaswamy}, ``{Fully-integrated non-magnetic
  non-reciprocal components based on linear periodically-time-varying
  circuits},'' in \emph{2017 IEEE 17th Topical Meeting on Silicon Monolithic
  Integrated Circuits in RF Systems (SiRF)}, Jan 2017, pp. 111--114.

\bibitem{campbel_2019}
C.~F. {Campbell}, ``{Demonstration of a Sequentially Switched Delay Line (SSDL)
  Circulator with SAW Filter Delay Elements},'' in \emph{2019 IEEE MTT-S
  International Microwave Symposium (IMS)}, June 2019.

\bibitem{harish_2017}
T.~{Dinc}, M.~{Tymchenko}, A.~{Nagulu}, D.~{Sounas}, A.~{Alù}, and
  H.~{Krishnaswamy}, ``{Synchronized conductivity modulation to realize
  broadband lossless magnetic-free non-reciprocity},'' \emph{Nature
  {Communication}}, vol.~8, pp. 1--9, Oct 2017.

\end{thebibliography}

\end{document}